\documentclass[pra,floatfix,showpacs,twocolumn,superscriptaddress]{revtex4}
\usepackage{amstext}
\usepackage{amsmath}
\usepackage{amssymb}
\usepackage{graphicx}
\usepackage{latexsym}
\usepackage{bm}
\usepackage{amsfonts}
\providecommand{\U}[1]{\protect\rule{.1in}{.1in}}
\begin{document}
\title{Quantum Theory of Transmission Line Resonator-Assisted Cooling of a
Micromechanical Resonator}
\author{Yong Li}
\affiliation{Department of Physics, University of Basel, Klingelbergstrasse 82,
4056 Basel, Switzerland}
\author{Ying-Dan Wang}
\affiliation{Department of Physics, University of Basel, Klingelbergstrasse 82,
4056 Basel, Switzerland}
\affiliation{NTT Basic Research Laboratories, NTT Corporation, Atsugi-shi,
Kanagawa 243-0198, Japan}
\author{Fei Xue}
\affiliation{CREST, Japan Science and Technology Agency (JST), Kawaguchi,
Saitama 332-0012, Japan} \affiliation{Frontier Research System, The Institute
of Physical and Chemical Research (RIKEN), Wako-shi, Saitama 351-0198, Japan}
\affiliation{Department of Electrical Engineering, Technion, Haifa 32000,
Israel}

\author{C. Bruder}
\affiliation{Department of Physics, University of Basel, Klingelbergstrasse 82,
4056 Basel, Switzerland}
\date{\today }

\begin{abstract}
We propose a quantum description of the cooling of a micromechanical flexural
oscillator by a one-dimensional transmission line resonator via a force that
resembles cavity radiation pressure. The mechanical oscillator is capacitively
coupled to the central conductor of the transmission line resonator. At the
optimal point, the micromechanical oscillator can be cooled close to the ground
state, and the cooling can be measured by homodyne detection of the output
microwave signal.
\end{abstract}

\pacs{85.85.+j, 45.80.+r, 03.67.Mn, 42.50.Lc}

\maketitle

\section{Introduction}

Micro- and nano-mechanical resonators have been an interesting research topic
due to their broad application in technology and fundamental physics
\cite{Braginsky}. This includes studies of ultrahigh precision displacement
detection \cite{ultrahigh detection}, mass detection \cite{Ekinci:2004},
gravitational-wave detectors \cite{gravitational-wave01,gravitational-wave02},
and attempts to observe quantum behavior of mechanical motion
\cite{quantum-mechanical01,quantum-mechanical02,quantum-mechanical03,quantum-mechanical04,quantum-mechanical05}.
Many of the applications are fundamentally limited by thermal fluctuations, and
in order to reduce their effects, it is desirable to cool the mechanical
oscillators. Recently, various schemes like the laser sideband cooling schemes
developed for trapped ions and atoms \cite{Wineland:1979}, have been proposed
for significantly cooling a mechanical resonator (MR) coupled to a Cooper-pair
box \cite{Martin:2004,Zhang:2005,Hauss:2008,Jaehne:2008}, a flux qubit
\cite{Wang:2007,You:2008}, a superconducting single-electron transistor
\cite{SSET}, quantum dots \cite{Wilson-Rae:2004}, trapped ions
\cite{Tian:2004}, and optical cavities
\cite{Metzger:2004,Gigan:2006,Arcizet:2006,Kleckner:2006,Schliesser:2006,Corbitt:2007,Thompson:2007,Schliesser:2008,Mancini:1998,Vitali:2002,Nori:2007,Paternostro:2006,Wilson-Rae:2007,Marquardt:2007,Bhattacharya:2007,Genes:2008,Dantan:2007,Kippenberg2007,Marquardt2008}%
. On the experimental side, optomechanical cooling schemes have been
shown to be promising
\cite{Metzger:2004,Gigan:2006,Arcizet:2006,Kleckner:2006,Schliesser:2006,Corbitt:2007,Thompson:2007,Schliesser:2008}:
the MR was cooled to ultra-low temperatures via either photothermal
forces or radiation pressure by coupling it to a driven cavity.
There are two main optomechanical cooling schemes. The first one
involves an active feedback loop
\cite{Kleckner:2006,Mancini:1998,Vitali:2002}, and the second one
works via passive back-action cooling (also called self-cooling)
\cite{Gigan:2006,Arcizet:2006,Schliesser:2006,Corbitt:2007}. A fully
quantum-mechanical description of cavity-assisted cooling schemes
for optomechanical systems has been given in
Refs.~\cite{Paternostro:2006,Wilson-Rae:2007,Marquardt:2007,Bhattacharya:2007,Genes:2008,Dantan:2007}
(for a review, see \cite{Kippenberg2007,Marquardt2008}).
Ground-state cooling of a mechanical resonator via passive cooling
schemes based on radiation pressure has also been investigated
theoretically
\cite{Wilson-Rae:2007,Marquardt:2007,Genes:2008,Dantan:2007}.

Recently, other optomechanical-like cooling schemes have been proposed to
replace the optical cavity by a radio-frequency (RF) circuit
\cite{Wineland:2006,Brown:2007} or a one-dimensional transmission line
resonator (TLR) \cite{Xue:2007b}. However, the theoretical understanding of the
cooling schemes via a RF circuit in Refs. \cite{Wineland:2006,Brown:2007} or
via a TLR in Ref. \cite{Xue:2007b} is based on a classical description of the
motion of the MR. A quantum-mechanical description of the motion of the MR, in
a similar system consisting of a mechanical resonator capacitively coupled to a
superconducting coplanar waveguide (which is an example of a TLR), was
discussed recently in Ref. \cite{Vitali:2007b}, which focused on studying the
entanglement between the MR and the TLR without considering the cooling of MR.
Most recently, Teufel \textit{et al.} \cite{Teufel:2008} considered the cooling
of a MR by applying directly the theoretical analysis of the cavity-assisted
back-action cooling scheme \cite{Marquardt:2007} to a superconducting microwave
resonator. They also presented experimental data about the cooling effect on
the MR due to the microwave radiation field. The quantum-mechanical description
of TLR-assisted cooling of a MR has also been investigated in Ref.
\cite{Blencowe:2007} via embedding a SQUID \cite{Buks:2007}, which allows to
control the coupling strength between MR and TLR by controlling the flux
through the SQUID.

There are some practical advantages \cite{Regal:2008,Marquardt:2008}
in the microwave TLR schemes. The TLR is realized in a thin on-chip
superconducting film and is easily pre-cooled by standard dilution
refrigeration techniques. It is ready to be integrated with quantum
circuits containing Josephson junctions which may offer a sensitive
measurement and a connection with quantum information processing. In
addition, the size of the mechanical resonator could be much smaller
than the wavelength of the radiation in the TLR, unlike in optical
cavity experiments that work with reflection.

In this paper, we present a quantum-mechanical description and use it to
investigate the motion of the MR when it is coupled capacitively to a driven
TLR as in Ref.~\cite{Xue:2007b} where the calculation was carried out in a
semi-classical framework. The Hamiltonian of our TLR-assisted model is also
studied in Refs.~\cite{Vitali:2007b,Teufel:2008}, and is very similar to that
of a MR coupled to a driven optical cavity via radiation pressure coupling
\cite{Paternostro:2006,Wilson-Rae:2007,Marquardt:2007,Bhattacharya:2007,Genes:2008,Dantan:2007,Kippenberg2007,Marquardt2008}.
We study the TLR-assisted passive back-action cooling of a MR in detail by
using a quantum Langevin description (without taking into account quantum
entanglement \cite{Vitali:2007b} between the MR and the TLR). One of the main
results of our work is to show that the MR can be cooled close to its ground
state using realistic parameters: final effective mean phonon numbers below $1$
can be reached assuming an initial temperature of $10$ mK which can be achieved
using a dilution refrigerator. We discuss in detail how such a ground state
cooling of the MR can be obtained for all kind of parameter choices in
practice.

\section{Model and Hamiltonian}

\begin{figure}[th]
\includegraphics[width=7cm]{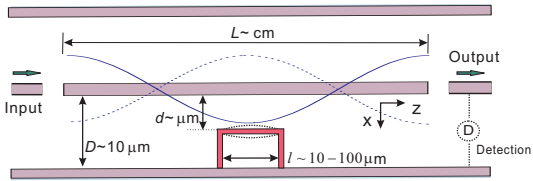}
\caption{(Color online)Schematic diagram of a mechanical resonator (MR) located
at the center of a one-dimensional transmission-line resonator (TLR). The
external microwave drive field enters from the left and drives the TLR. The
signal at the output on the right end can be used to measure the motion of the
MR via homodyne detection
\cite{Paternostro:2006,Teufel:2008,Giovannetti:2001,Regal:2008}.} \label{fig1}
\end{figure}

The system that we consider is shown schematically in Fig.~\ref{fig1}: a MR is
fixed on both ends (or a cantilever fixed on one end) located at the center of
the TLR and is coupled capacitively to the central conductor of the TLR
\cite{Blais:2004}. We restrict the description to the fundamental flexural mode
of oscillation of the MR which is modeled as a harmonic oscillator of frequency
$\omega_{b}$ and effective mass $m$. The TLR is driven by an external microwave
at a frequency $\omega_{d}$ and can be modeled as a single mode LC resonator
with frequency $\omega_{a}^{\prime}=1/\sqrt{L_{a}C_{a}}$ (the second mode of
the TLR \cite{second harmonic}), where $L_{a}$ is the inductance and $C_{a}$
the capacitance of the TLR.
%

The Hamiltonian of the system reads
\begin{align}
H  &  =\hbar\omega_{a}^{\prime}a^{\dagger}a+\left(  \frac{p^{2}}{2m}
+\frac{m\omega_{b}^{2}}{2}x^{2}\right) \nonumber\\
&  +\frac{C_{g}(x)}{2}V^{2}+\hbar(\varepsilon a^{\dagger}e^{-i\omega_{d}
t}+\varepsilon^{\ast}ae^{i\omega_{d}t}). \label{ham0}
\end{align}
The first line describes the free Hamiltonian of the TLR and the MR,
respectively, with lowering (rising) operator of the TLR mode $a$ ($a^{\dag}
$), and the position (momentum) operator of the MR $x$ ($p$) which satisfy
$[a,a^{\dag}]=1$ and $[x,p]=i\hbar$. The first term in the second line is the
capacitive coupling between the TLR and the MR. Actually, it describes the
capacitive energy between them. The MR and the TLR are assumed to form a
capacitor with the capacitance $C_{g}(x)\approx C_{g}^{0}(1-x/d)$ (for small
displacement) depending on the position of the MR along the $x$-direction ($d$
is the initial equilibrium distance without the coupling and $C_{g}^{0}$ the
corresponding initial capacitance; typically $d\sim1$ $\mu$m
\cite{Regal:2008}). The capacitor is assumed to be placed in the center of the
structure, i.e., its voltage is given by the antinode voltage of the second
mode: $V=V_{rms}(a^{\dagger}+a)$ (where
$V_{rms}=\sqrt{\hbar\omega_{a}^{\prime}/C_{a}}$ is the rms voltage
\cite{Blais:2004}), since the length of the MR is usually much shorter than
that of the TLR: $L\sim$ cm $\gg l\sim10-100$ $\mu$m. The last term in
Eq.~(\ref{ham0}) describes the input driving of the TLR by an external
microwave field with the coupling strength
\cite{Gigan:2006,Genes:2008,Blais:2004} $|\varepsilon|=\sqrt{2\kappa
P/\hbar\omega_{a}^{\prime}}$, where $\kappa$ is the decay rate of the TLR, $P$
is the input external microwave drive power. Here, the non-rotating wave terms
like $ae^{-i\omega_{d}t}$ and $a^{\dagger}e^{i\omega_{d}t}$ have been ignored
since we keep $|\varepsilon|\ll\omega_{a}^{\prime}\sim\omega_{d}$.

Usually, the fundamental oscillation frequency is of the order of $2\pi\times$
($10^{3}$ - $10^{6}$)~Hz for micromechanical resonators and $2\pi\times$
($10^{7}$ - $10^{9}$)~Hz for nanomechanical resonators; the TLR frequency can
be made to be of the order of $2\pi\times10$~GHz. Here we will focus on the
case of a micro-MR for which the condition $\omega_{b}$ $\ll$ $\omega
_{a}^{\prime}$ is satisfied. In the interaction picture with respect to
$\hbar\omega_{d}a^{\dag}a$ and neglecting the rapidly-oscillating terms, the
Hamiltonian reads
\begin{align}
H_{I}  &  =\hbar\Delta_{0}a^{\dagger}a+\left(  \frac{p^{2}}{2m}+\frac
{m\omega_{b}^{2}}{2}x^{2}\right) \nonumber\\
&  -\frac{\hbar g_{0}}{2}(2a^{\dagger}a+1)x+\hbar(\varepsilon a^{\dagger
}+\varepsilon^{\ast}a)
\end{align}
where $g_{0}:=C_{g}^{0}{V_{rms}^{2}}/(\hbar d)$ is a real coupling constant;
$\Delta_{0}=\omega_{a}-\omega_{d}$ is the detuning, and
$\omega_{a}=\omega_{a}^{\prime}+C_{g}^{0}V_{rms}^{2}/\hbar$ is the modified
frequency of the TLR shifted by the coupling between TLR and MR.

This Hamiltonian resembles that used in cavity-assisted cooling schemes
\cite{Paternostro:2006,Wilson-Rae:2007,Marquardt:2007,Bhattacharya:2007,Genes:2008,Dantan:2007}%
. This suggests that the capacitive-coupling scheme in a microwave TLR can be
used to cool the MR like in the case of radiation-pressure cooling in an
optical cavity.

\section{Quantum Langevin equations and final mean phonon number}

The dynamics is also determined by fluctuation-dissipation processes that
affect both the TLR and the mechanical mode. They are taken into account in a
fully consistent way by the quantum Langevin equations \cite{Gardiner:book}:
\begin{subequations}
\label{nonlinlang1}
\begin{align}
\dot{x}  &  =p/m ,\\
\dot{p}  &  =-m\omega_{b}^{2}x-\gamma_{b}p+\frac{\hbar g_{0}}{2}(2a^{\dagger
}a+1)+\xi ,\\
\dot{a}  &  =-(\kappa+i\Delta_{0})a+ig_{0}ax+\varepsilon+\sqrt{2\kappa} a_{in}.
\label{a}
\end{align}
\end{subequations}
Here $a_{in}$ ($a_{in}^{\dagger}$) is the noise operator due to the external
microwave drive, and $\xi(t)$ denotes the quantum Brownian force that the
resonator is subject to. They satisfy \cite{Gardiner:book}
\begin{align}
&  \left\langle a_{in}(t)a_{in}^{\dagger}(t^{\prime})\right\rangle
=(N+1)\delta(t-t^{\prime}),\\
&  \left\langle \xi(t)\xi(t^{\prime})\right\rangle =\frac{\hbar\gamma_{b}
m}{2\pi}\int\mathrm{d}\omega\mathrm{e}^{-i\omega(t-t^{\prime})}\omega
(1+\coth\frac{\hbar\omega}{2k_{B}T}),
\end{align}
where $N=1/[\exp(\hbar\omega_{a}/k_{B}T)-1]$ is the mean number of
thermal microwave photons of the TLR, $k_{B}$ is the Boltzmann
constant, $T$ the temperature of the environment, and $\gamma_{b}$
the damping rate of the MR. For simplicity, we have assumed that
both the bath correlated to the external microwave drive field and
the one connected to the MR have the same temperature
\cite{Vitali:2007b}. We now perform a similar calculation as that
given in
Refs.~\cite{Paternostro:2006,Genes:2008,Dantan:2007,Vitali:2007}.
The steady-state solution of the quantum Langevin
equations~(\ref{nonlinlang1}) can be obtained by first replacing the
operators by their average and then setting $\mathrm{d}\left\langle
...\right\rangle /\mathrm{d}t=0$. Hence we can get the steady-state
values as
\begin{equation}
\left\langle p\right\rangle =0,\quad\left\langle x\right\rangle =\frac{\hbar
g_{0}\left(  \left\vert \left\langle a\right\rangle \right\vert ^{2}+\frac
{1}{2}\right)  }{m\omega_{b}^{2}},\quad\left\langle a\right\rangle
=\frac{\varepsilon}{\kappa+i\Delta}, \label{nonlinlang3-1}
\end{equation}
where $\Delta=\Delta_{0}-g_{0}\left\langle x\right\rangle $ is the effective
detuning. In Eq.~(\ref{nonlinlang3-1}), we can also take $|\langle
a\rangle|^{2}+\frac{1}{2}\simeq|\langle a\rangle|^{2}$, since we will focus on
the case $\left\vert \left\langle a\right\rangle \right\vert \gg1$ which can be
achieved by controlling the input power of the external microwave drive.

Rewriting each operator as a $c$-number steady-state value plus an
additional fluctuation operator, and neglecting the nonlinear terms
(since we have chosen $\vert \langle a\rangle \vert \gg1$), we
obtain a set of linear quantum Langevin equations (see
Eq.~(\ref{nonlinlang42})) and then solve for the spectrum of the
position and momentum of the MR as in
Refs.~\cite{Paternostro:2006,Genes:2008,Dantan:2007,Vitali:2007},
see Appendix A.

Using the fluctuation spectra of the MR as given in
Eqs.~(\ref{spectr-x},\ref{spectr-p}), we can define the final mean phonon
number in the steady state \cite{Genes:2008} as
\begin{equation}
n_{b}^{\text{f}}=\frac{\left\langle \delta p^{2}\right\rangle }{2\hbar
m\omega_{b}}+\frac{m\omega_{b}}{2\hbar}\left\langle \delta x^{2}\right\rangle
-\frac{1}{2}, \label{n_eff}
\end{equation}
where the variances of position and momentum are
\begin{equation}
\left\langle \delta r^{2}\right\rangle =\frac{1}{2\pi}\int_{-\infty}^{+\infty
}S_{r}(\omega)\mathrm{d}\omega,\quad(r=x,p). \label{integral}
\end{equation}
This allows us to define the effective temperature $T_{\text{eff}}$ as
\begin{equation}
T_{\text{eff}}=\frac{\hbar\omega_{b}}{k_{B}}\ln^{-1}(\frac{1}{n_{b}^{\text{f}
}}+1). \label{T_eff}
\end{equation}

In the next section, we will consider the cooling of the MR by
discussing its final effective mean phonon number (or equivalently,
its effective temperature) in detail.

\section{Cooling of the MR}

The final effective mean phonon number of the MR can be calculated directly by
evaluating the integral in Eq.~(\ref{integral}) numerically and using
Eq.~(\ref{n_eff}). Alternatively, instead of being evaluated directly,
Eq.~(\ref{integral}) can also be evaluated analytically using the approximation
scheme described in the following.

The effective mechanical damping rate $\gamma_{b}^{\text{eff}}(\omega
)=\gamma_{b}+\gamma_{ca}(\omega)$ can be significantly increased, $\left\vert
\gamma_{b}^{\text{eff}}(\omega)\right\vert \gg\gamma_{b}$, when $\left\vert
g_{0}\left\langle a\right\rangle \right\vert $ is very large, see
Eq.~(\ref{gamma_ca}). Let us consider the most interesting regime when the
significantly increased effective mechanical damping rate is less than the
mechanical frequency: $|\gamma_{b}^{\text{eff}}(\omega)|<\omega_{b}$, (that is,
the effective quality factor
$Q_{b}^{\text{eff}}=\omega_{b}/|\gamma_{b}^{\text{eff}}(\omega)|>1$), and also
less than the decay rate of TLR: $|\gamma_{b}^{\text{eff}}(\omega)|<\kappa$
\cite{Dantan:2007,Pinard:2005,Wilson-Rae:2007,Marquardt:2007}. In this regime,
the effective frequency is unchanged $\omega_{b}^{\text{eff}}(\omega
)\simeq\omega_{b}$ \cite{Genes:2008,note} according to Eq.~(\ref{omega_eff}),
and the effective susceptibility is peaked around the points $\omega=\pm
\omega_{b}^{\text{eff}}(\omega)\simeq\pm\omega_{b}$. Then one can get an
approximate expression for the variance
\begin{equation}
\left\langle \delta x^{2}\right\rangle \approx\frac{S_{th}^{\prime}(\omega
_{b})+S_{ca}^{\prime}(\omega_{b})}{2m^{2}\omega_{b}^{2}{\left\vert \gamma
_{b}^{\text{eff}}(\omega_{b})\right\vert }}, \label{approximate-x}
\end{equation}
where the effective thermal noise spectrum $S_{th}^{\prime}(\omega)$ and the
induced noise spectrum $S_{ca}^{\prime}(\omega)$ are the symmetrized parts of
$S_{th}(\omega)$ and $S_{ca}(\omega)$, respectively:
\begin{align}
S_{th}^{\prime}(\omega)  &  =\hbar\gamma_{b}m\omega\coth\frac{\hbar\omega
}{2k_{B}T},\\
S_{ca}^{\prime}(\omega)  &  =\left(  2N+1\right)  \hbar m\frac{\kappa
^{2}+\Delta^{2}+\omega^{2}}{2\Delta}\gamma_{ca}(\omega).
\end{align}
Similarly, one can obtain
\begin{equation}
\left\langle \delta p^{2}\right\rangle =\left(  m\omega_{b}\right)
^{2}\left\langle \delta x^{2}\right\rangle . \label{approximate-p}
\end{equation}
%
\begin{figure}[th]
\includegraphics[width=7 cm]{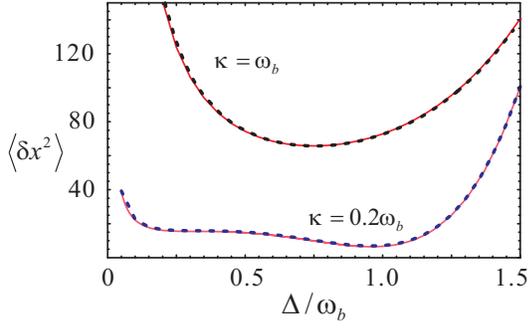}\caption{(Color
online) Variance of position $\left\langle \delta x^{2}\right\rangle $ in units
of $\hbar/m\omega_{b}$ as a function of effective detuning $\Delta$. The dashed
lines are obtained by numerically evaluating the integral in
Eq.~(\ref{integral}), the solid lines by using the approximate expressions
Eqs.~(\ref{approximate-x},\ref{approximate-p}). Here, $T=6\times10^{3}
\hbar\omega_{b}/k_{B}$, $g_{0}=3\times10^{-5}\omega_{b}\sqrt{m\omega_{b}
/\hbar}$, $\omega_{a}=2\times10^{4}\omega_{b}$, $\varepsilon=2.5\times
10^{3}\omega_{b}$, $\gamma_{b}=0.25\times10^{-4}\omega_{b}$, and
$\kappa=\omega_{b}$ (upper lines) or $\kappa=0.2\omega_{b}$ (lower lines).}
\label{fig2}
\end{figure}

Figure~\ref{fig2} shows the variance of position $\langle \delta x^{2}\rangle$
as a function of the effective detuning $\Delta$. The dashed lines correspond
to a numerical evaluation of the integral in Eq.~(\ref{integral}). The solid
lines describe the approximate results obtained through
Eq.~(\ref{approximate-x}) which can be seen to agree perfectly with the exact
numerical evaluation. We checked that this is also the case for the variance of
the momentum $\langle \delta p^{2} \rangle $.

In Eq.~(\ref{approximate-x}), the induced noise spectrum $S_{ca}
^{\prime}(\omega_b)$ increases (heats) the motion of the MR. On the
other hand, when the effective damping rate is enhanced:
$|\gamma_{b}^{\text{eff}} (\omega_b)|>\gamma_{b}$, the mechanical
motion will reduce, that means cooling. Mathematically, the cooling
effect would dominate the heating effect when the effective damping
rate is sufficiently increased. Actually, this is right when the
significantly increased effective damping rate is positive for
positive detuning. However, it is not the case when
$\gamma_{b}^{\text{eff} } (\omega_{b})$ is negative and $\left\vert
\gamma_{b}^{\text{eff}}(\omega _{b})\right\vert \gg\gamma_{b}$ (for
negative detuning $\Delta<0$). That is because the stability
conditions, derived using Ref.~\cite{Hurwitz:book}, are satisfied
only for positive detuning
\cite{Paternostro:2006,Vitali:2007,Genes:2008b}. In fact, a negative
effective damping means the amplitude motion of the MR will be
amplified which will lead to an instability
\cite{Bennett2006,Armour2007,Ludwig2008}.

In what follows, we will focus on the case of positive detuning $\Delta>0$.
According to Eqs.~(\ref{n_eff},\ref{approximate-x},\ref{approximate-p}), one
has
\begin{equation}
n_{b}^{\text{f}}=\frac{\gamma_{b}n_{b} +\gamma_{ca}(\omega_{b})n_{ca}}
{\gamma_{b}+\gamma_{ca}(\omega_{b})},\label{nb-eff}
\end{equation}
where
\begin{equation}
n_{b}\equiv\frac{S_{th}^{\prime}(\omega_{b})}{2\hbar m\gamma_{b}\omega_{b}
}-\frac{1} {2}\equiv\frac{1}{\exp(\hbar\omega_{b}/k_{B}T)-1}\label{nb}
\end{equation}
is the initial mean thermal excitation phonon number of the MR;
\begin{equation}
n_{ca}\equiv\frac{2N+1}{4\omega_{b}\Delta}\left(  \kappa^{2}+\Delta^{2}
+\omega_{b}^{2}\right)  -\frac{1}{2}\label{nca}
\end{equation}
is the induced mean phonon number due to the capacitive coupling between the MR
and the TLR.
%
\begin{figure}[th]
\includegraphics[width=7cm]{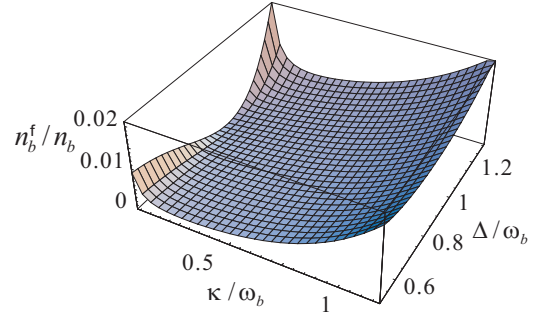}\caption{(Color online)
Final mean phonon number in the steady state $n_{b}^{\text{f}}$ vs. effective
detuning $\Delta$ and decay rate $\kappa$ of the TLR ($T=3\times10^{4}
\hbar\omega_{b}/k_{B}$, for the other parameters see Fig.~\ref{fig2}).}
\label{fig3}
\end{figure}

As discussed above, the significant reduced value of
$n_{b}^{\text{f}}$ in Eq.~(\ref{nb-eff}) can only be obtained when
the additional damping rate $\gamma_{ca}(\omega_{b})$ (or effective
damping rate $\gamma_{b}^{\text{eff} }(\omega_{b})$) is positive and
much larger than the original one (but still less than the decay
rate of TLR and less than the frequency of the MR as discussed
before):
\begin{equation}
\gamma_{b}\ll\gamma_{ca}(\omega_{b})<\left( \omega_{b}, \kappa\right) .
\label{condition-1}
\end{equation}
This can be satisfied by enhancing the value of $\left\vert g_{0}\left\langle
a\right\rangle \right\vert $, that is, by controlling the capacitive coupling
strength $g_{0}$ and increasing the external microwave drive power $P$ to make
$\left\vert \varepsilon\right\vert $ large (equivalently, $\left\vert
\left\langle a\right\rangle \right\vert $ will be large). In practice, the
capacitive coupling strength $g_{0}$ would be limited by the realistic system,
and the external microwave drive strength $\varepsilon$ would also be limited
according to the validity of the rotating wave approach as we mentioned before.
Here we put the length of the MR $l$ as large as $10-100$ $\mu$m and the
distance between the MR and TLR $d$ as small as $1$ $\mu$m (see
Fig.~\ref{fig1}) in order to get a large capacitance $C_{g}^{0}$ which will
lead to large $g_{0}$, and fix $\left\vert \varepsilon\right\vert = \omega
_{a}/8$ for all the numerical calculations. Then for a significantly-increased
effective damping rate, the final mean phonon number reduces to
\begin{equation}
n_{b}^{\text{f}}=\frac{\gamma_{b}}{\gamma_{ca}(\omega_{b})}n_{b}
+n_{ca}.\label{nb-eff2}
\end{equation}
In order to get the ground state cooling, that is, $n_{b}^{\text{f}}\ll 1$,
both $n_{ca}$ and $n_{b}\gamma_{b}/\gamma_{ca}(\omega_{b})$ should be much less
than $1$. Especially, if the $\gamma_{ca}(\omega_{b})$ is significantly
increased enough to make
\begin{equation}
\gamma_{b}n_{b}\ll\gamma_{ca}(\omega_{b})n_{ca}, \label{limit-condition}
\end{equation}
then $n_{b}^{\text{f}}$ approaches the limit $n_{ca}$:
\begin{equation}
n_{b}^{\text{f}} \rightarrow n_{ca}.
\end{equation}

Now we discuss the possible minimal value of $n_{ca}$ by discussing all kinds
of parameters, e.g., $\kappa$, $\Delta$, and $N$. From Eq.~(\ref{nca}), it is
obvious that the optimal value of $\kappa$ satisfies the high-quality cavity
limit
\begin{equation}
\kappa^{2}\ll\omega_{b}^{2},
\end{equation}
and the optimal detuning satisfies
$\Delta=\sqrt{\omega_{b}^{2}+\kappa^{2}}\approx\omega_{b}$. Then the
corresponding induced mean phonon number $n_{ca}$ is
\begin{equation}
n_{ca}\approx N+\frac{\kappa^{2}}{4\omega_{b}^{2}}.\label{general limit}
\end{equation}

The optimal $N$ needs a sufficiently low initial temperature of the bath which
is limited in practice to the experimental dilute refrigerator temperatures.
For the superconducting TLR scheme, its microwave frequency is of the order of
$2\pi\times10^{10}$ Hz. For the initial temperature $T\gtrsim 1$ K,
$k_{B}T\gtrsim\hbar\omega_{a}$ and $N\gtrsim1$, the ground-state cooling of the
MR is not possible. Therefore, initial temperature less than $100$ mK are
required to achieve ground-state cooling.

Our result on the limiting value in Eq.~(\ref{general limit}) is consistent
with that in other optical schemes except the limit of initial temperatures. In
the optical cavity case $\hbar\omega_{a}\gg k_{B}T$ (even at room temperature)
and therefore $N\simeq0$, the optimal value of $n_{ca}$ becomes
\begin{equation}
n_{ca}\approx\frac{\kappa^{2}}{4\omega_{b}^{2}}, \label{n_eff2}
\end{equation}
which is just the case of resolved sideband cooling as discussed in the
optomechanical cooling schemes
\cite{Gigan:2006,Arcizet:2006,Schliesser:2006,Corbitt:2007,Schliesser:2008}. We
would like to point out that these references also mention another cooling
limit: the Doppler cooling limit, which is realized in our system when
$N\simeq0$, $\Delta=\sqrt{\omega_{b}^{2}+\kappa^{2}} $, and
$\kappa^{2}\gg\omega_{b}^{2}$ in Eq.~(\ref{nca}):
\begin{equation}
n_{ca}\approx\frac{\kappa}{2\omega_{b}}>1. \label{dopper-limit}
\end{equation}
On the other hand, if $N\gg\kappa^{2}/4\omega_{b}^{2}$, the induced mean phonon
number $n_{ca}$ in Eq.~(\ref{general limit}) becomes $n_{ca}\rightarrow N$. In
the classical limit when the initial temperature is so high that $N\approx
k_{B}T/(\hbar\omega_{a})\gg1$, one has
\begin{equation}
\frac{T_{\text{eff}}}{T}=\frac{n_{b}^{\text{f}}}{n_{b}}\approx\frac{n_{ca}
}{n_{b}}=\frac{\omega_{a}}{\omega_{b}}, \label{classical limit}
\end{equation}
which is also given in Ref.~\cite{Nori:2007}. The Doppler cooling limit in
Eq.~(\ref{dopper-limit}) and the classical cooling limit in Eq.~(\ref{classical
limit}) preclude ground state cooling. We will focus on the resolved sideband
cooling in this paper.

The final mean phonon number $n_{b}^{\text{f}}$ is plotted as a function of the
effective detuning $\Delta$ and the decay rate $\kappa$ of the TLR in
Fig.~\ref{fig3}. It is clear that one can obtain a significant suppression of
the mechanical motion of the MR in the positive detuning range $\Delta
\simeq\omega_{b}$. The optimal cooling is obtained for $\kappa^{2}\ll
\omega_{b}^{2}$, which agrees with both the above analysis and that in other
treatments of radiation-pressure cooling
\cite{Gigan:2006,Arcizet:2006,Schliesser:2006,Corbitt:2007,Xue:2007b}.

Physically, as discussed in the back-action optomechanical cooling
schemes in optical cavities
(Refs.~\cite{Wilson-Rae:2007,Marquardt:2007,Genes:2008,Dantan:2007,Genes:2008b}),
the external driving microwave is scattered by the ``TLR + MR"
system mostly to the first Stokes sideband ($\omega_{d} -
\omega_{b}$) and the first anti-Stokes sideband ($\omega_{d} +
\omega_{b}$). The generation of an anti-Stokes photon will cool the
MR by taking away a phonon of the MR. On the contrary, the
generation of a Stokes photon will heat the MR by creating a phonon.
When the effective detuning $\Delta>0$, the microwave field of the
TLR (with the frequency $\omega_{a} \equiv\omega_{d}+\Delta_{0}$
$\approx\omega_{d} + \Delta$) interacts with the first anti-Stokes
sideband ($\omega_{d} + \omega_{b}$) more than it interacts with the
first Stokes sideband ($\omega_{d} - \omega_{b}$), and cooling will
occur. This is the physical reason why the positive effective
detuning ($\Delta>0$) will lead to cooling. In the high-quality
cavity limit $\kappa< \omega_{b}$, the anti-Stokes (Stokes) sideband
is resolved, and the corresponding cooling (heating) process is
prominent. Especially, for the optimal effective detuning
$\Delta\approx\omega_{b}$, the frequency of the TLR is resonant with
that of the anti-Stokes sideband, which will apparently lead to
optimal cooling. This physical discussion is consistent with the
calculation presented above.

Figure~\ref{fig3} suggests there is a finite optimal value of $\kappa$ for a
fixed effective detuning. That is because one should have both a small value of
$n_{ca}$ and a large effective damping rate $\gamma _{ca}(\omega_{b} )$ in
order to get strong cooling: $\kappa$ should not be too large, since $n_{ca}$
depends somewhat on the value of $\kappa^{2} /\omega _{b}^{2}$; $\kappa$ should
not be too small, since $\gamma_{ca} (\omega _{b})\rightarrow$ $0$ when
$\kappa\rightarrow0$ \cite{note3}. In the cooling process, the thermal energy
of the MR is mainly first transferred to the TLR, and then leaks out of the TLR
through the bath coupled to the TLR. When the decay rate of the TLR is too
small: $\kappa\rightarrow 0$, the energy leakage out of the TLR is too weak,
and one could not obtain a strong cooling.
%
\begin{figure}[th]
\includegraphics[width=6cm]{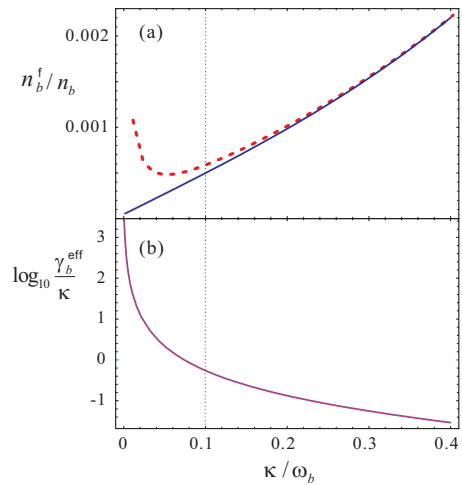}\caption{(Color online)
(a) Optimal final mean phonon number obtained by numerically evaluating the
integral in Eq.~(\ref{integral}) (dashed line) or using the approximate
expression in Eqs.~(\ref{approximate-x},\ref{approximate-p}) (solid line) as a
function of $\kappa$ at the optimal effective detuning $\Delta=\omega_{b}$. (b)
Logarithm of the ratio of the corresponding effective damping rate
$\gamma_{b}^{\text{eff}}(\omega_{b})$ to $\kappa$ as a function of $\kappa$.
Here, $T=3\times10^{3}\hbar\omega_{b}/k_{B}$. For the other parameters see
Fig.~\ref{fig2}.} \label{new-fig4}
\end{figure}

We would like to emphasize that the results shown in Fig.~\ref{fig3} are based
on the approximate expressions Eqs.~(\ref{approximate-x},\ref{approximate-p}),
where the condition of the so-called weak-coupling limit
\cite{Marquardt:2007,Teufel:2008} has been assumed, that is, the effective
damping rate of the MR should be always less than the decay rate of the cavity
and less than the frequency of the MR
$|\gamma_{b}^{\text{eff}}(\omega_{b})|<\kappa,\omega_{b}$. Normally the
weak-coupling is satisfied but not in some special cases. In
Fig.~\ref{new-fig4}(b), the weak-coupling condition is violated when
$\kappa/\omega_{b}<0.1$ at the optimal effective detuning $\Delta=\omega_{b}$.
Beyond the weak coupling limit, Fig.~\ref{new-fig4}(a) shows that the
approximate treatment through Eqs.~(\ref{approximate-x},\ref{approximate-p})
ceases to be valid. Then one should discuss the cooling, e.g., effective mean
phonon number in Eq.~(\ref{n_eff}), by using the numerical evaluation of the
integral in Eq.~(\ref{integral}). But going beyond the weak-coupling limit, the
contribution from the position variance is not equivalent to that from the
momentum variance any more \cite{Genes:2008}. In other words, the energy
equipartition is not satisfied. That means it is hard to define an effective
temperature since it is not in a strict thermal state.

According to the above analysis, both the high-quality cavity and weak-coupling
limit should be satisfied, so the optimal decay rate of the TLR is better taken
to be $\kappa\approx 0.1\omega_{b}$ for the typical parameters in
Fig.~\ref{new-fig4}. The weak-coupling condition depends only weakly on the
initial temperature $T$ and the original damping rate of the MR $\gamma_{b}$ in
the cooling process. In what follows, we will consider the optimal decay rate
at $\kappa\approx 0.1\omega_{b}$ for different parameters $T$ and $\gamma_{b}$,
for which the weak coupling limit is always satisfied.

\begin{figure}[th]
\includegraphics[width=7cm]{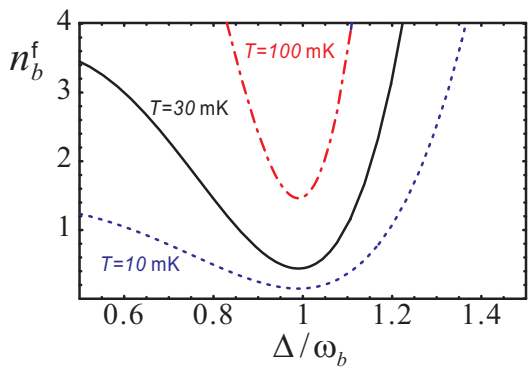}
\caption{(Color online) The final mean phonon number vs. effective detuning
$\Delta$ for three initial temperatures: $T=10$ mK (dotted lines), $T=30$ mK
(solid lines), $T=100$ mK (dot-dashed lines). Here, $\kappa=0.1\omega_{b}$,
$m=1.5\times10^{-13}$ kg, $\omega_{b}=4$ MHz, $\gamma_{b}=0.25\times
10^{-4}\omega_{b}$ (equivalently, $Q_{b}\equiv\omega_{b}/\gamma_{b}
=4\times10^{4}$). For the other parameters see Fig.~\ref{fig2} .}
\label{new-fig5}
\end{figure}

In Fig.~\ref{new-fig5}, the ratio of final effective temperature
$T_{\text{eff}}$ to bath temperature $T$ is plotted as a function of the
effective detuning $\Delta$ for the optimal $\kappa\approx 0.1\omega_{b}$.
Apparently, here the weak coupling limit is satisfied (according to the above
analysis in Fig.~\ref{new-fig4}).
For initial temperatures $T=100$, $30$, and $10$~mK, the corresponding initial
mean phonon numbers are $n_{b}=1/[\exp(\hbar\omega_{b}/k_{B}T)-1]$ $\simeq
k_{B}T/\hbar\omega_{b}$ $\simeq3300$, $980$, and $330$, with the final mean
phonon number $n_{b}^{\text{f}}\approx1.6$, $0.5$, and $0.16$, respectively. It
is obvious that a significant cooling of the MR is obtained and lower initial
temperatures will generally lead to better cooling. For an initial temperature
$T=10$~mK, which can be realized experimentally by using a dilution
refrigerator, the MR (with the frequency $\omega_{b}\sim4$~MHz) can be cooled
close to the ground state since the final mean phonon number
$n_{b}^{\text{f}}\approx0.16<1$.
%

\begin{figure}[th]
\includegraphics[width=7cm]{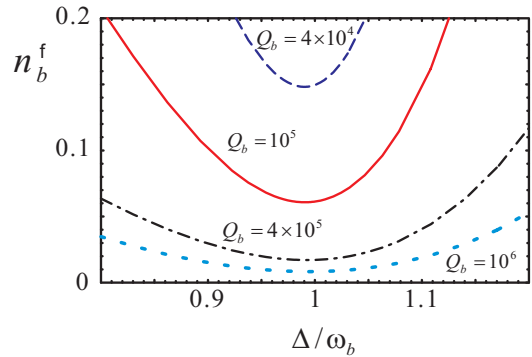}\caption{(Color online)
The final effective mean phonon number $n_{b}^{\text{f}}$ is plotted as a
function of $\Delta$ for different quality factors of the MR: $Q_{b}
=4\times10^{4}$, $10^{5}$, $4\times10^{5}$, $10^{6}$ (from up to down), the
corresponding damping rate $\gamma_{b}=100$, $40$, $10$, $4$ Hz. Here, $T=10$
mK. For the other parameters see Fig.~\ref{new-fig5}.} \label{new-fig6}
\end{figure}

For an initial temperature of $T=10$~mK as in Fig.~\ref{new-fig5}, the final
mean phonon number would in principle be $n_ca$ (in Eq.~(\ref{nca})), which is
much less than that obtained in Fig.~\ref{new-fig5}:
$n_{ca}=N+\kappa^{2}/4\omega_{b}^{2}$ $\approx0.0025$ $\ll n_{b}^{\text{f}
}\approx0.16$. This is because $n_{b}^{\text{f}}$ $\rightarrow n_{ca}$ only
when the condition in Eq.~(\ref{limit-condition}) is satisfied. Unfortunately,
it is not the case for the parameters in Fig.~\ref{new-fig5}. A possible way to
approach this condition is to increase the quality factor of the MR. In
Fig.~\ref{new-fig6}, the final effective mean phonon number is plotted as a
function of the effective detuning $\Delta$ for different quality factors of MR
$Q_{b}$ ($\equiv\omega_{b}/n_{b}$): $Q_{b}=4\times 10^{4}$ (typically, see Ref.
\cite{Teufel:2008}); $Q_{b}=10^{5}$, $4\times 10^{5}$, $10^{6}$ (expected in
the near future).
The corresponding minimal $n_{b}^{\text{f}}\approx 0.16$, $0.06$, $0.02$,
$0.01$. One can find the cooling is better for a higher quality factor of the
MR.
%
%

The cooling discussed above can be measured by a homodyne detection method like
that given in the scheme of cavity-assisted radiation-pressure cooling of a MR
\cite{Paternostro:2006,Teufel:2008,Giovannetti:2001,Regal:2008}. The motion of
the MR can be detected by monitoring the output microwave signal (e.g., the
field phase quadrature) of the TLR (as seen in Fig.~\ref{fig1}) since the
measurement of the output spectrum corresponds to a faithful measurement of the
MR motion \cite{Regal:2008}.

\section{Conclusion}

We have found that a MR with frequency $\omega_{b}\sim2\pi\times 10^{6}$ Hz can
be cooled close to its ground state when it is coupled to a typical TLR
($\omega_{a}\sim2\pi\times10^{10}$ Hz). Actually, by considering the optimal
parameters in this scheme, that is, assuming the high-quality cavity limit
($\kappa^{2}\ll\omega_{b}^{2}$), a positive optimal effective detuning
($\Delta\approx\omega_{b}$), a low initial temperature (e.g., $T=10$ mK in
order that $N\approx10^{-27}\simeq0$), a high quality factor of the MR
($Q_{b}\equiv\omega_{b}/\gamma_{b}\gtrsim10^{4}$), and both strong external
input microwave drive power $P$ and strong capacitive coupling strength $g_{0}$
to get the significantly increased positive effective damping rate
($\gamma_{b}\ll\gamma_{b}^{\text{eff}}(\omega_{b})\approx\gamma_{ca}(\omega_{b})$),
we find that resolved sideband cooling of the MR occurs. The possible minimal
value of the final effective phonon number could approach the induced mean
phonon number: $n_{b}^{\text{f}}\rightarrow n_{ca}$
$\approx\kappa^{2}/4\omega_{b}^{2}\ll1$. Moreover, one should also consider the
condition of weak-coupling limit, that is, the significantly increased
effective damping rate $\gamma_{b}^{\text{eff}} (\omega_{b})$ should be less
than both $\omega_{b}$ and $\kappa$ (in the high-quality cavity limit, one only
needs $\gamma_{b}^{\text{eff}}(\omega_{b})<\kappa $). This condition requires
that $\kappa$ must not be too small though lower $\kappa$ will lead to lower
$n_{ca}$. 
As shown in Fig.~\ref{fig2} and its discussion, there will be an optimal range
of $\kappa$. For the typical parameters in this cooling scheme, we take
$\kappa\sim0.1\omega_{b}$ (though this is not the optimal result in general).
We find that a MR with $\omega_{b}=4$ MHz can be cooled close to its ground
state with the final effective mean phonon number in the steady state:
$n_{b}^{\text{f}}\approx0.16$ (for a typical quality factor
$Q_{b}=4\times10^{4}$) or $n_{b}^{\text{f}}\approx0.01$ (for a high quality
factor $Q_{b}=10^{6}$) by using the resolved sideband cooling scheme when it is
coupled to a driven TLR (with the frequency $\omega_{a}=8\times10^{10}$ Hz).

We would like to stress the condition in Eq.~(\ref{limit-condition}), which can
lead to $n_{b}^{\text{f}}\rightarrow n_{ca}$. As pointed out in the discussion
of Figs.~(\ref{new-fig5}, \ref{new-fig6}), this is not always satisfied. A
possible way to approach this condition is to increase the quality factor of
the MR. For example, if the quality factor of the MR is high enough (e.g.,
$Q_{b}>10^{7}$), 
one would have the optimal $n_{b}^{\text{f}}\rightarrow n_{ca}\approx\kappa
^{2}/4\omega_{b}^{2}=0.0025$, for which the MR is cooled much closer to the
ground state.

The back-action self-cooling scheme presented here is similar to the
optical-cavity-assisted cooling scheme \cite{Wilson-Rae:2007,Marquardt:2007}.
In both cases, the MR can be cooled close to its ground state using resolved
sideband cooling which is possible in the limit of a high-quality cavity. But
it seems that this limit (e.g., $\kappa=0.1\omega_{b}$, for
$\omega_{b}=4\times10^{6}$ Hz) is easier to reach in the microwave TLR than
that in the optical cavity. In the case of a TLR (typically
$\omega_{a}=8\times10^{10}$ Hz), quality factors of
$Q_{a}=\omega_{a}/\kappa=2\times10^{5}$ have been seen in experiments
\cite{Regal:2008,Day:2003}. However, in the case of an optical cavity
($\omega_{a}\sim 10^{15}$ Hz), the corresponding quality factor should be
$Q_{a}=\omega_{a}/\kappa \sim2.5\times10^{9}$, which is hard to achieve since
the typical cavity quality factor is $Q_{a}\sim10^{7-8}$ \cite{Blais:2004}.

To conclude, we have studied the self-cooling of a mechanical resonator that is
capacitively coupled to a transmission-line resonator. The discussion was based
on a linearized quantum Langevin equation. The cooling method presented here is
similar to the self-cooling of a MR coupled to an optical cavity by radiation
pressure. By using the optimal parameters discussed above, the MR can be cooled
close to its ground state in the high-quality cavity and weak-coupling limit.

\begin{acknowledgments}

We would like thank C.B. Doiron and I. Wilson-Rae for helpful
discussions. This work was supported by the Swiss NSF, the NCCR
Nanoscience, the EC IST-FET project EuroSQUIP, and partially supported
by the NSFC through Grant No.~10574133. Y.D.W. also acknowledges
support by the JSPS KAKENHI No.~18201018 and MEXT-KAKENHI
No.~18001002. F.X.  was supported in part at the Technion by an Aly
Kaufman Fellowship.
\end{acknowledgments}

\appendix

\section{Equivalence to time-dependent second-order perturbation theory}

\label{appendix}

The linearized quantum Langevin equations read
\begin{subequations}
\label{nonlinlang42}
\begin{align}
\delta\dot{x}  &  =\delta p/m,\\
\delta\dot{p}  &  =-m\omega_{b}^{2}\delta x-\gamma_{b}\delta p+\hbar
g_{0}(\delta a^{\dagger}\left\langle a\right\rangle +h.c.)+\xi,\\
\delta\dot{a}  &  =-(\kappa+i\Delta)\delta a+ig_{0}\left\langle a\right\rangle
\delta x+\sqrt{2\kappa}a_{in}. \label{nonlinlang42-d}
\end{align}
\end{subequations}

To solve these equations, we define the Fourier transform for an operator $u$
($u= \delta a$, $\delta x$, $\delta p$, $a_{in}$, $\xi$)
\begin{equation}
u(t):=\frac{1}{\sqrt{2\pi}}\int_{-\infty}^{+\infty}\mathrm{e}^{i\omega
t}\tilde{u}(\omega)\mathrm{d}\omega,
\end{equation}
and for its Hermitian conjugate $u^{\dag}$ (if any)
\begin{equation}
u^{\dag}(t):=\frac{1}{\sqrt{2\pi}}\int_{-\infty}^{+\infty}\mathrm{e}^{-i\omega
t}\tilde{u}^{\dag}(\omega)\mathrm{d}\omega,
\end{equation}
which lead to
\begin{align}
&  \left\langle \tilde{a}_{in}(\Omega)\tilde{a}_{in}^{\dagger}(\omega
)\right\rangle =(N+1)\delta(\Omega-\omega),\\
&  \left\langle \tilde{\xi}(\Omega)\tilde{\xi}(\omega)\right\rangle
=\hbar\gamma_{b}m\omega(1+\coth\frac{\hbar\omega}{2k_{B}T})\delta
(\Omega+\omega).
\end{align}

After solving the linear quantum Langevin equations in the frequency domain, we
obtain
\begin{subequations}
\label{nonlinlang51}
\begin{align}
\delta\tilde{x}(\omega)  &  =\frac{C^{\ast}(-\omega)\tilde{a}_{in}
+C(\omega)\tilde{a}_{in}^{\dagger}+\left[  (\kappa+i\omega)^{2}+\Delta
^{2}\right]  \tilde{\xi}}{B(\omega)},\label{x}\\
\delta\tilde{p}(\omega)  &  =i\omega m\delta\tilde{x}(\omega), \label{p}
\end{align}
\end{subequations}
where $B(\omega)=m(\omega_{b}^{2}-\omega^{2}+i\gamma_{b}\omega)\left[
(\kappa+i\omega)^{2}+\Delta^{2}\right]  -2\hbar|g_{0}\langle a\rangle
|^{2}\Delta$, and $C(\omega)=\hbar\sqrt{2\kappa}g_{0}\langle a\rangle[\kappa +
i(\omega+\Delta)]$.

To calculate the effective temperature of the MR, we define the fluctuation
spectra of position and momentum
\cite{Paternostro:2006,Genes:2008,Gardiner:book} of the MR, which are given by
the following correlation function:
\begin{align}
S_{x}(\omega)  &  =\int_{-\infty}^{+\infty}\mathrm{e}^{-i\omega\tau
}\left\langle \delta x(t+\tau)\delta x(t)\right\rangle _{s}d\tau,\\
S_{p}(\omega)  &  =\int_{-\infty}^{+\infty}\mathrm{e}^{-i\omega\tau
}\left\langle \delta p(t+\tau)\delta p(t)\right\rangle _{s}d\tau.
\end{align}

Here, $\langle...\rangle_{s}$ denotes the steady-state average. Equivalently,
$S_{x,p}(\omega)$ can also be defined as

\begin{equation}
\left\langle \delta\tilde{r}(\Omega)\delta\tilde{r}(\omega)\right\rangle
_{s}:=S_{r}(\omega)\delta(\Omega+\omega),\ \ (r=x,p).
\end{equation}

According to Eqs.~(\ref{nonlinlang51}), the spectra of the MR can be written as
\begin{align}
S_{x}(\omega)  &  \equiv\left\vert \chi_{\text{eff}}(\omega)\right\vert
^{2}\left[  S_{th}(\omega)+S_{ca}(\omega)\right]  ,\label{spectr-x}\\
S_{p}(\omega)  &  =\left(  \omega m\right)  ^{2}S_{x}(\omega), \label{spectr-p}
\end{align}
where
\begin{equation}
S_{th}(\omega)=\hbar\gamma_{b}m\omega [1+\coth(\hbar\omega/2k_{B}T)]
\end{equation}
is the thermal noise spectrum due to the Brownian motion of the MR; and
\begin{align}
S_{ca}(\omega)  &  =\frac{(N+1)|C(\omega)|^{2}+N|C(-\omega)|^{2}}
{|(\kappa+i\omega)^{2}+\Delta^{2}|^{2}}\nonumber\\
&  =2\hbar^{2}|g_{0}\langle a\rangle|^{2}\kappa\frac{\left(  2N+1\right) \left(
\kappa^{2}+\Delta^{2}+\omega^{2}\right)  +2\omega\Delta}
{|(\kappa+i\omega)^{2}+\Delta^{2}|^{2}} \label{s-ca}
\end{align}
is the induced noise spectrum due to the capacitive coupling to the driven TLR
The effective susceptibility is defined as $\chi_{\text{eff}}(\omega)=\left[
(\kappa+i\omega)^{2}+\Delta^{2}\right]  /B(\omega)$ and can be simplified to
\begin{equation}
\chi_{\text{eff}}(\omega)\equiv\frac{1}{m\left[  \left(  \omega_{b}
^{\text{eff}}(\omega)\right)  ^{2}-\omega^{2}+i\omega\gamma_{b}^{\text{eff}
}(\omega)\right]  },
\end{equation}
where the effective frequency of the MR is

\begin{align}
\omega_{b}^{\text{eff}}(\omega)  &  =\sqrt{\omega_{b}^{2}-\frac{2\hbar
\left\vert g_{0}\left\langle a\right\rangle \right\vert ^{2}\Delta\left(
\kappa^{2}-\omega^{2}+\Delta^{2}\right)  }{m\left\vert (\kappa+i\omega
)^{2}+\Delta^{2}\right\vert ^{2}}}\label{omega_eff}\\
&  \equiv\sqrt{\omega_{b}^{2}-\frac{\left(  \kappa^{2}-\omega^{2}+\Delta
^{2}\right)  \gamma_{ca}(\omega)}{2\kappa}},\nonumber
\end{align}
and the effective damping rate $\gamma_{b}^{\text{eff}}(\omega)=\gamma
_{b}+\gamma_{ca}(\omega)$ with the additional term
\begin{equation}
\gamma_{ca}(\omega)=\frac{4\hbar\left\vert g_{0}\left\langle a\right\rangle
\right\vert ^{2}\kappa\Delta}{m\left\vert (\kappa+i\omega)^{2}+\Delta
^{2}\right\vert ^{2}} \label{gamma_ca}
\end{equation}
resulting from the capacitive coupling.

According to the definition of the additional damping rate in Eq.
(\ref{gamma_ca}), the induced noise spectrum $S_{ca}(\omega)$ in Eq.
(\ref{s-ca}) can also expressed as
\begin{equation}
S_{ca}(\omega)=m\hbar\left[  \left(  2N+1\right)  \frac{\kappa^{2}+\Delta
^{2}+\omega^{2}}{2\Delta}+\omega\right]  \gamma_{ca}(\omega). \label{s-ca2}
\end{equation}


\begin{thebibliography}{99}
\bibitem {second harmonic}Here we assumed the mechanical resonator to be
fabricated at the center of the TLR. As a result, the qubit is coupled to the
second mode of the TLR, which as an antinode of the voltage in its center as
given in Ref. \cite{Blais:2004}.
\bibitem {note}Strictly speaking, $\omega_{b}^{\text{eff}}(\omega)\simeq
\omega_{b}$ is not always satisfied for arbitrary parameters, but this is the
case for the parameters considered in the following discussion.
\bibitem{note3} It is right that $\gamma_{ca} (\omega_{b})\rightarrow$ $0$
when $\kappa\rightarrow0$ except the scope nearly-close to $\Delta \equiv
\omega_b $ where $\gamma_{ca} (\omega_{b})\rightarrow$ $\infty$ when
$\kappa\rightarrow0$. But we will show in Fig.~\ref{new-fig4} that some
correction should be done since it may be beyond the weak-coupling limit.
\end{thebibliography}
\end{document}